\documentclass[aps,pr,reprint,groupedaddress,superscriptaddress]{revtex4-2}
\usepackage{amsmath,amssymb,amsfonts,latexsym,cancel} 
\usepackage[none]{hyphenat}
\usepackage{graphics}
\usepackage{natbib}
\usepackage[latin1]{inputenc}
\usepackage[pdftex]{graphicx} 
\usepackage[pdftex,colorlinks=true,linkcolor=blue,citecolor=blue,urlcolor=blue]{hyperref}
\usepackage{xcolor}
\begin{document}
\title{Imaging cloaked objects: diffraction tomography of realistic invisibility devices}
\author{Francisco J. D\'iaz-Fern\'andez}
\email{fradafer@ntc.upv.es}
\affiliation{Nanophotonics Technology Center, Universitat Polit\`{e}cnica de Val\`{e}ncia, 46022, Valencia, Spain}
\author{Javier Mart\'i}
\affiliation{Nanophotonics Technology Center, Universitat Polit\`{e}cnica de Val\`{e}ncia, 46022, Valencia, Spain}
\affiliation{DAS Photonics S.L., 46022, Valencia, Spain}
\author{Carlos Garc\'ia-Meca}
\email{cgarcia@dasphotonics.com}
\affiliation{DAS Photonics S.L., 46022, Valencia, Spain}
\begin{abstract}
Invisibility cloaks have become one of the most outstanding developments among the wide range of applications in the field of metamaterials. So far, most efforts in invisibility science have been devoted to achieving practically realizable cloak designs and to improving the effectiveness of these devices in reducing their scattering cross-section (SCS), a scalar quantity accounting for the total electromagnetic energy scattered by an object. In contrast, little attention has been paid to the opposite side of the technology: the development of more efficient techniques for the detection of invisibility devices. For instance, the SCS ignores the phase change introduced by the cloak, as well as the angular dependence of the incident and scattered waves. Here, we propose to take advantage of the smarter way in which diffraction tomography processes all this overlooked information to improve the efficiency in unveiling the presence of invisibility devices. We show that this approach not only results in a considerable sensitivity enhancement in the detection of different kinds of cloaks based on both scattering cancellation and transformation optics, but also enables us to even obtain images depicting the approximate shape and size of the cloak. As a side result, we find that scattering cancellation coatings may be used to enhance the contrast of tomographic images of closely-packed particles. Our work could be easily extended to the detection of sound cloaks.  
\end{abstract}

\maketitle
\section{Introduction}
\begin{figure}[ht]
	\centering
	\includegraphics{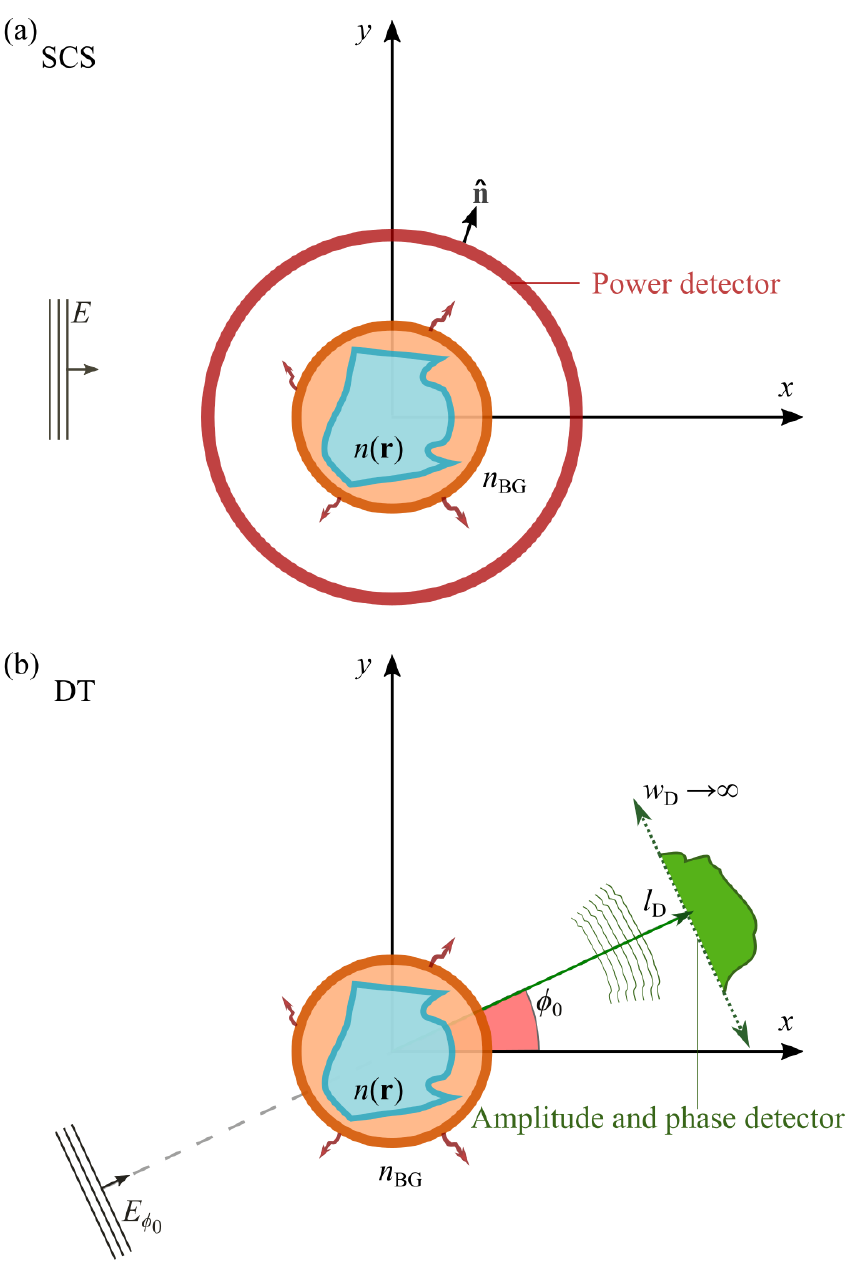}
	\caption{Scheme of the difference between the SCS and DT detection techniques. (a) The SCS is a scalar quantity related to the total scattered power for a given incident direction. The phase of the scattered field is not taken into account. (b) DT uses different angles of illumination to obtain a RI map of the object from the space-dependent phase and amplitude of the scattered field.}
	\label{Fig00}
\end{figure}

Invisibility has been one of the most challenging effects pursued by humankind for centuries. The possibility of hiding objects to the naked eye has recently leaped from science fiction to a feasible reality thanks to the advent of metamaterials \cite{Engheta2006}. Among a wide range of applications that arises through the exploitation of this kind of materials not available in nature, invisibility cloaks are one of the most high-impact developments \cite{Alu2005,Pendry2006}. The impressive ability of these devices to hide objects by reducing the scattering they produce has even been experimentally demonstrated~\cite{Schurig2006, Rainwater2012}, boosting the impact of this field of study over the last two decades. Different approaches to the achievement of invisibility have followed in different fields besides optics. For instance, a variety of cloaking devices have been experimentally demonstrated also in acoustics, thermodynamics and mechanics \cite{Kadic2015}. 
However, cloaking devices are not perfect. It has been shown that outside the design frequency, realistic cloaks may become significantly visible~\cite{Monticone2013}. Conversely, as technology improves, detecting a cloak at the design frequency might be a challenging task. In most previous studies, the scattering cross section (SCS) has been used as an indicator of the effectiveness of realistic invisibility cloaks \cite{Chen2012, Monticone2016, Monticone2013}. The SCS is a measurement that estimates the total energy scattered by an object (see Appendix~\ref{AP.SCS}). Therefore, the use of this scalar value has the disadvantage that it does not take into account the spatial distribution of the scattered field nor the phase changes produced by the cloak and, in the case of devices without rotational symmetry, only a single illumination direction is usually considered. It is thus reasonable to imagine that a suitable measurement and processing of these missing data would provide more information on the cloak, which, as predicted in \cite{Monticone2013}, should be more sensible to interferometric techniques than to the SCS. Actually, looking back at the origins of invisibility cloaks, we find that one of the first known invisibility devices was conceived as a material undetectable by tomographic techniques \cite{Greenleaf2003}, which aim at reconstructing the refractive index (RI) profile of an object by illuminating it from each possible direction and by smartly combining the complex amplitude (magnitude and phase) of the resulting scattered waves~\cite{Devaney1982}. Therefore, diffraction tomography (DT) is expected to detect invisibility devices more accurately, and even has the potential of revealing the shape of the cloak (see Fig.~\ref{Fig00}).

\begin{figure}[!ht]
	\centering
	\includegraphics[width=6cm]{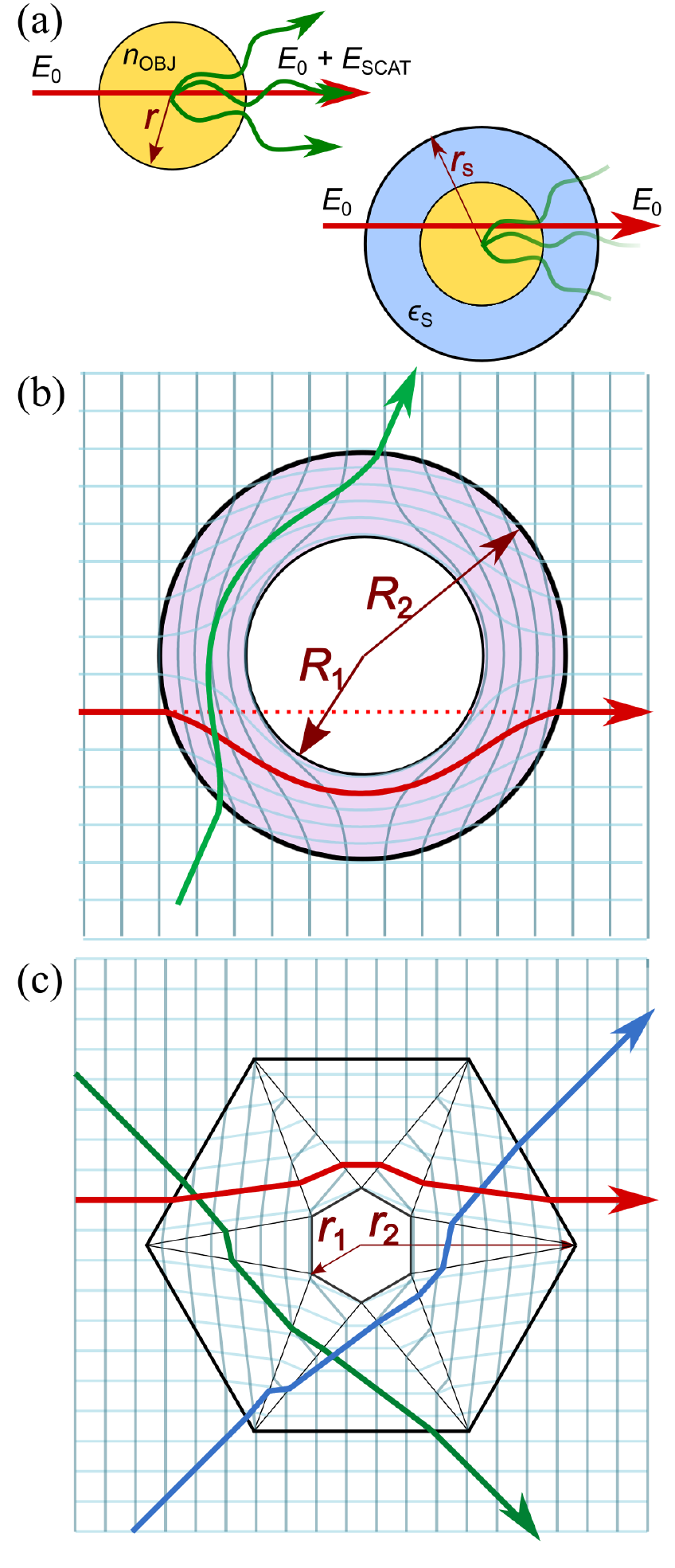}
	\caption{Studied invisibility cloaks and their working principle. (a) SC cloak (b) Blow-up-a-point TO cloak. (c) Polygonal TO cloak.}
	\label{Fig_cloaks}
\end{figure}

In this paper, we explore this possibility, keeping the spotlight on studying the behavior of passive cloaks under DT around their design frequency, at which the SCS measurement is less effective and may fail to unveil their presence. Particularly, we analyze three of the most representative kinds of invisibility devices: plasmonic cloaks based on scattering cancellation (SC), blow-up-a-point inhomogeneous cloaks based on transformation optics (TO), and homogeneous polygonal cloaks, also based on TO (see Fig.~\ref{Fig_cloaks}). 

In all cases, we consider realistic material implementations. Additionally, it is worth mentioning that DT is constrained by the diffraction limit, i.e., it is suited for imaging objects with dimensions larger than $\lambda/2$, where $\lambda$ is the wavelength of light, if both the forward and backward scattering are measured. If the latter is not considered, which is the common situation and the one we will consider in this work, the limit rises to $\lambda/\sqrt{2}$ \cite{Muller2015}, so we will restrict our study to this class of objects. Moreover, we will focus on two-dimensional (2D) problems for simplicity, although the extension to the three-dimensional case is straightforward. Likewise, we restrict ourselves to TE waves (electric field polarized along the $z$ component and magnetic field contained in the $XY$ plane), with analogous results for TM waves.

Finally, to implement the DT algorithms used in this paper, the steps described in~\cite{Muller2015,Muller2015b} have been coded in \textsc{Matlab}. These robust algorithms have been used, for example, to measure the refractive index of a single cell \cite{Schurmann2017}.

\section{Scattering cancellation}
The first approach to invisibility cloaking that we will study is scattering cancellation. This technique reduces the scattering produced by a dielectric object with a positive permittivity by covering it with a negative-permittivity coating [see Fig.~\ref{Fig_cloaks}(a)]~\cite{Chen2012}. SC has also been applied to cloak objects immersed in a uniform static magnetic field~\cite{Gomory2012} or in a diffusive light scattering medium~\cite{Schittny2014}. An interesting application of this type of cloak is the possibility of cloaking a sensor without affecting its capability to measure an incoming signal~\cite{Alu2009sensor}. Typically, the SC set-up consists of a core-shell spherical or cylindrical structure in which the oppositely signed permittivities of the inner and outer materials are designed to cancel the first-order scattered field at a given frequency (the design frequency)~\cite{Chen2012}. 

\begin{figure}
		\includegraphics{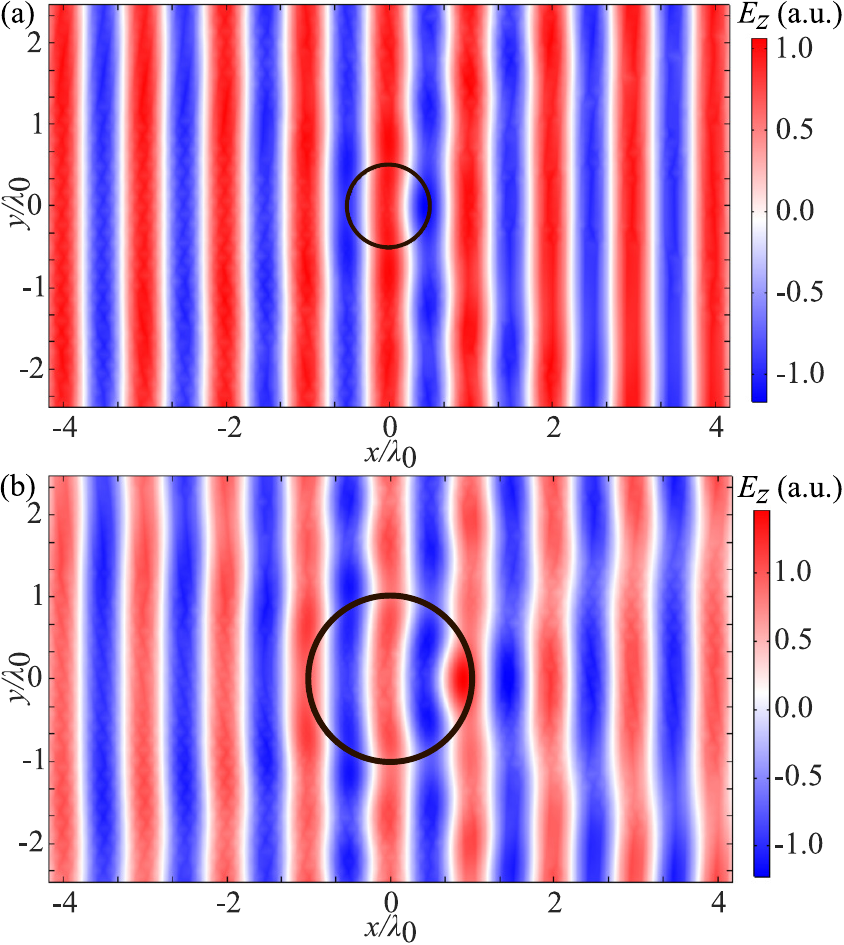}
		\caption{Simulated $z$ component of the electric field resulting from the illumination of a SC-cloaked dielectric cylinder of radius $r$ with an $x$-directed plane wave. (a) $r=\lambda_0/2$. (b) $r=\lambda_0$.}
		\label{FigOptEz}
	\end{figure}

Note that the SC effect is maximized for objects with electrically small dimensions~\cite{Chen2012}. Therefore, taking into account the diffraction limit of DT, we will analyze objects with longitudinal dimensions in the interval $[\lambda_0/\sqrt{2},2\lambda_0]$, where $\lambda_0$ is the free-space wavelength for which the cloaks are designed, with $f_0$ the corresponding frequency. Moreover, to satisfy the Born and Rytov approximations usually employed in DT~\cite{Muller2015,Chen1998}, the maximum contrast between the object RI and that of the background is fixed to a 5\%. In particular, we apply the SC technique to cloak dielectric cylinders of radii $\lambda_0/2$ and $\lambda_0$, with a RI $n_{\rm{OBJ} }=1.05$, and immersed in a vacuum background with RI $n_{\rm{BG}}=1$. These cloaks have been optimized to minimize the SCS following the procedure described in \cite{Silveirinha2007} (see Appendix \ref{AP.SCS}). This yields optimum metallic shells with a permittivity $\epsilon_{\rm{S}} = -2.7$ and an outer radius $r_{\rm{S}} =0.508 \lambda$ for the $r = \lambda_0/2$ cylinder and $r_{\rm{S}} = 1.0205 \lambda_0$ for the $r = \lambda_0$ cylinder.	 As mentioned above, the electric field is assumed to be polarized along the $z$ component (parallel to the cylinder axis). Figure~\ref{FigOptEz} shows the distribution of this field for both studied SC cloaks under plane wave incidence.             
	
To quantify the effectiveness of the cloak in terms of the SCS reduction it attains, we define the relative observability as:
\begin{equation}
O_{\rm SCS} = \frac{\rm SCS_C}{\rm SCS_{NC}},
\end{equation}
where $\rm SCS_C$ and $\rm SCS_{NC}$ are the SCS of the object with and without cloak, respectively. That is, the lower the value of $O_{\rm SCS}$, the higher the efficiency of the cloak. In the case of the $r = \lambda_0/2$ cylinder, a value of $O_{\rm SCS} = 0.7$ is achieved, while for the $r = \lambda_0$ cylinder, $O_{\rm SCS} = 0.66$.  

On the other hand, to obtain tomographic images (called \textit{tomograms}) of all the cloaks studied in this work, the scattered field is recorded on a detector line (this data set is called a \textit{projection}) placed at a distance $l_{\rm{D}}$ from the set-up center, as shown in Fig. \ref{Fig00}(b). In the case of the SC cylinders, we set $l_{\rm{D}}=10 \lambda_0$. Ideally, the detector line should be infinite to capture the complete forward scattered wave. In practice, in the simulations we use a finite line length $w_{\rm D}$ such that the missed scattered field is negligible. In the case of the SC cylinders, $w_{\rm D} = 220 \lambda_0$. A tomogram is built upon the information provided by different projections corresponding to different values of the illumination angle $\phi_0$ [see Fig. \ref{Fig00}(b)]. To ensure a nearly perfect tomographic quality, we take 250 projections and, at least, record the field of each projection at 8 points per $\lambda$ (a total of 1760 points in this case). 

\begin{figure}[h!]
	\includegraphics{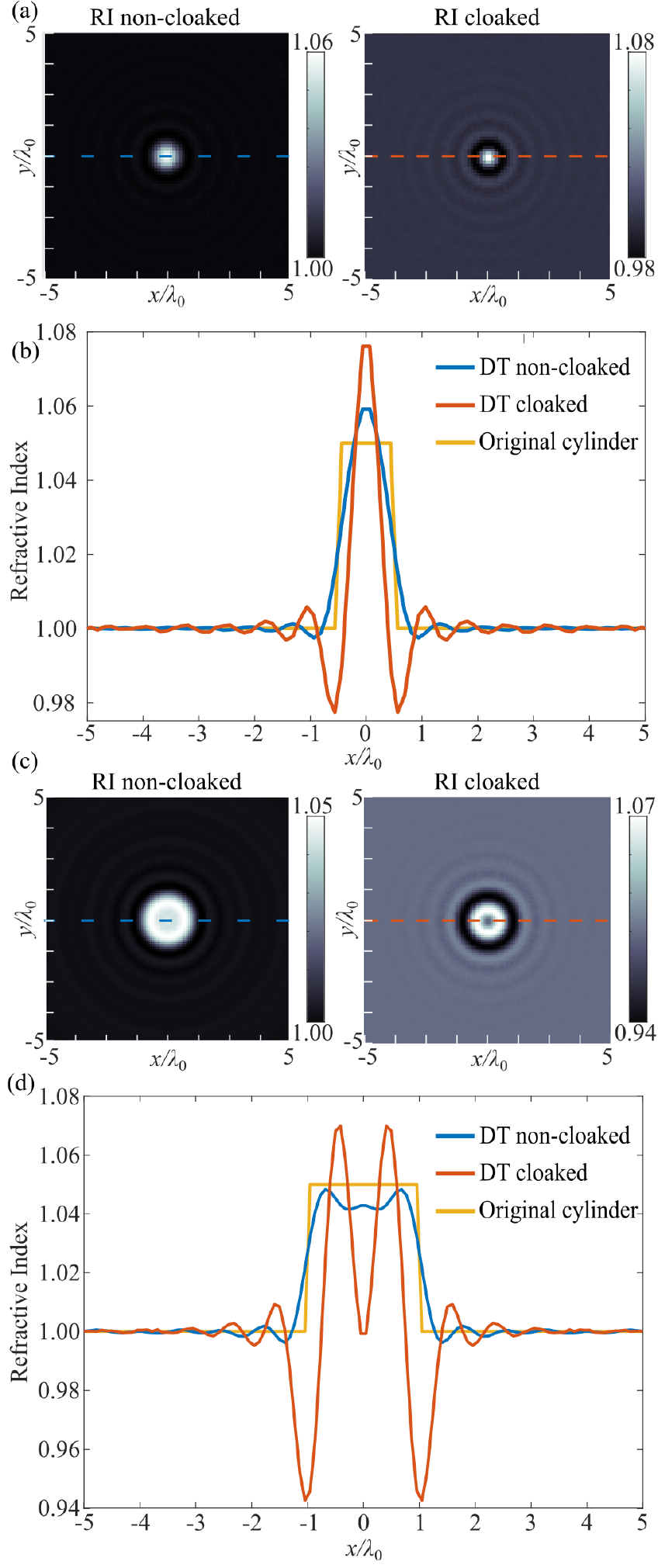}
	\caption{Linear tomography (Rytov approximation) of non-cloaked and cloaked dielectric cylinders of radius $r=\lambda_0/2$ (a,b) and $r=\lambda_0$ (c,d). Full tomograms are shown in panels (a) and (c). Panels (b) and (d) show a detail of the RI along the dashed lines depicted in (a) and (c).}
	\label{FigOptb}
\end{figure}

The obtained tomograms for the non-cloaked cylinders are shown in Fig. \ref{FigOptb}, yielding a good approximation to the original cylinder RI profiles. It is worth mentioning that the reconstructed RI has a smooth variation instead of the abrupt RI change of the original cylinders, as we can see in Fig.~\ref{FigOptb}(b) and Fig.~\ref{FigOptb}(d). These results are in line with those expected for the tomography of a centered cylinder \cite{Chen1998}. Notably, the RI recovered by the tomogram for the cloaked $r=\lambda_0/2$ configuration [Fig.~\ref{FigOptb}(b)] exhibits a higher amplitude variation than that of the non-cloaked configuration (0.1 with cloak, 0.06 without it). This increment of the RI range is more evident for the $r=\lambda_0$ cylinder, as can be seen in Fig.~\ref{FigOptb}(d). Additionally, in this case, the RI profile shows two lobes instead of one. Regardless, the tomograms clearly show the presence of an object in both cases.

\begin{figure}
 		\centering
 		\includegraphics{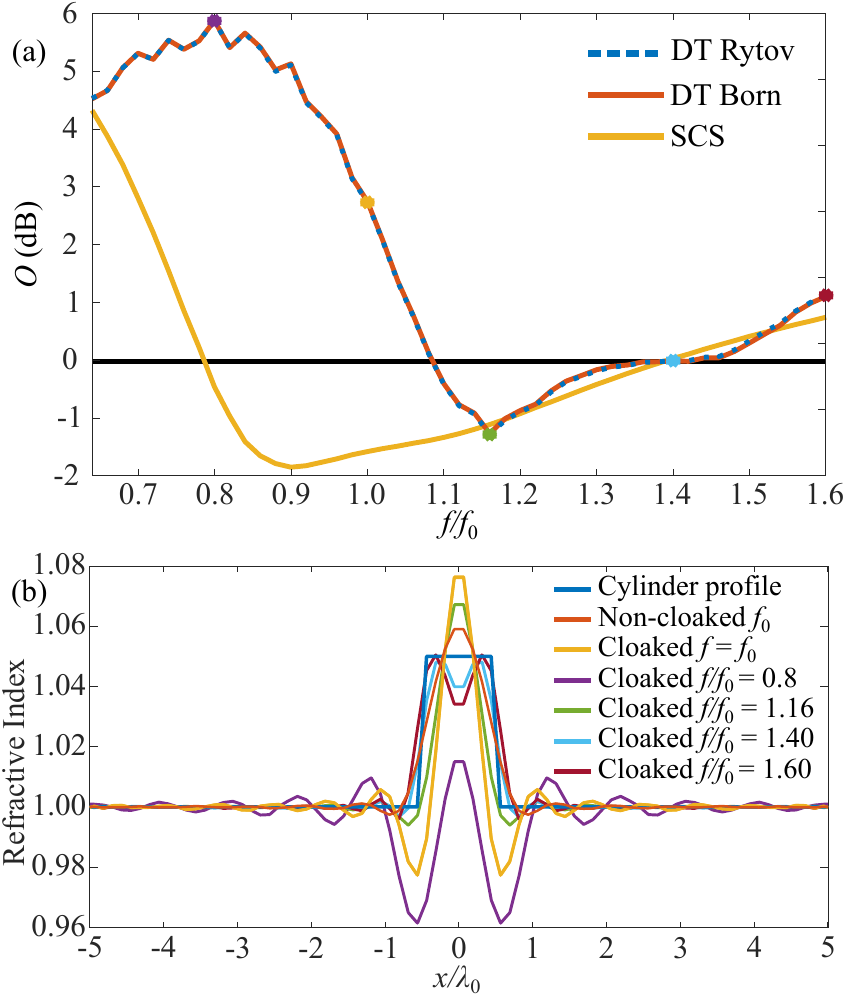}
 		\caption{(a) Comparison between the relative observabilities obtained via SCS and DT for an SC-cloaked cylinder with $r = \lambda_0/2$ and  $n_{\rm{OBJ}}=1.05$ as a function of frequency. (b) Corresponding RI along a line passing through the center of the cylinder as retrieved via DT for some selected frequencies [highlighted with colored points in (a)].}
 		\label{FigOptSC}
 \end{figure}

	Following the work in Ref.~\cite{Monticone2013}, it is worth analyzing the behavior of the cloak under DT at different frequencies. For that, the cloak was considered to be made of lossless silver~\cite{Monticone2013}, and was modeled by a Drude permittivity $\epsilon_{\rm{s}}=\epsilon_{\infty} - f_{p}^2 / f^2$, with a plasma frequency $f_p=2175$ THz and $\epsilon_{\infty}=5$ \cite{Johnson1972}. As a consequence, the ideal cloak permittivity is obtained at $f_0 =  783.8$ THz (i.e., $\lambda_0 = 382.7$ nm). To quantify the sensitivity of DT in the detection of a given cloak, we define the presence of an object according to its tomogram as:

 \begin{align}
 		p=\dfrac{\sum_{i=1}^{I} \sum_{j=1}^{J} \left| n_{\rm{DT}}(i,j) - n_{\rm{BG}}  \right|}{IJ},
 	\end{align}\label{presenceSC}
where $n_{\rm{DT}}(i,j)$ is the RI at the pixel $(i,j)$ of a tomogram of size $I \times J$. This is similar to the way in which the error of tomographic techniques is tested~\cite{Chen1998}. As in the case of the SCS, the effect of the cloak is quantified as the presence of the cloaked object divided by the presence of the non-cloaked one, i.e.:
\begin{equation}
O_{\rm DT} = \frac{p_{\rm C}}{p_{\rm NC}}.
\end{equation}
The detection ability of both measurement methods (SCS and DT) is shown in Fig.~\ref{FigOptSC}(a) for the $r=\lambda_0/2$ cylinder. The corresponding tomogram profiles along a line passing through the center of the cloak are shown in Fig.~\ref{FigOptSC}(b) for some selected frequencies. 
 
As expected, $O_{\rm SCS}$ shows a local minimum close to $f = f_0$, since $f_0$ is the design frequency. As noted in Ref.~\cite{Monticone2013}, there is a frequency range around $f_0$ for which the cloak reduces the scattering produced by the concealed object (that is, $O_{\rm SCS} < 1$ or, equivalently, $O_{\rm SCS} < 0$ dB), while the SCS is higher for the cloaked object in the rest of the spectrum ($f < 0.8 f_0$ and $f > 1.4 f_0$ in our case), as can be seen in Fig.~\ref{FigOptSC}(a). This corresponds to the values greater than 0 dB in this figure, which represent the frequencies at which the cloak not only fails at hiding the object, but enhances its presence, making it more detectable. Remarkably, in the eyes of DT, the presence of the cloaked system is almost always greater than that of the non-cloaked one in the studied band, even at the frequencies for which the cloak reduces the object SCS. Additionally, we have $O_{\rm DT} \geq O_{\rm SCS}$, so we conclude that DT will be more effective in discovering objects hidden by SC cloaks.

\begin{figure}
		\centering
		\includegraphics{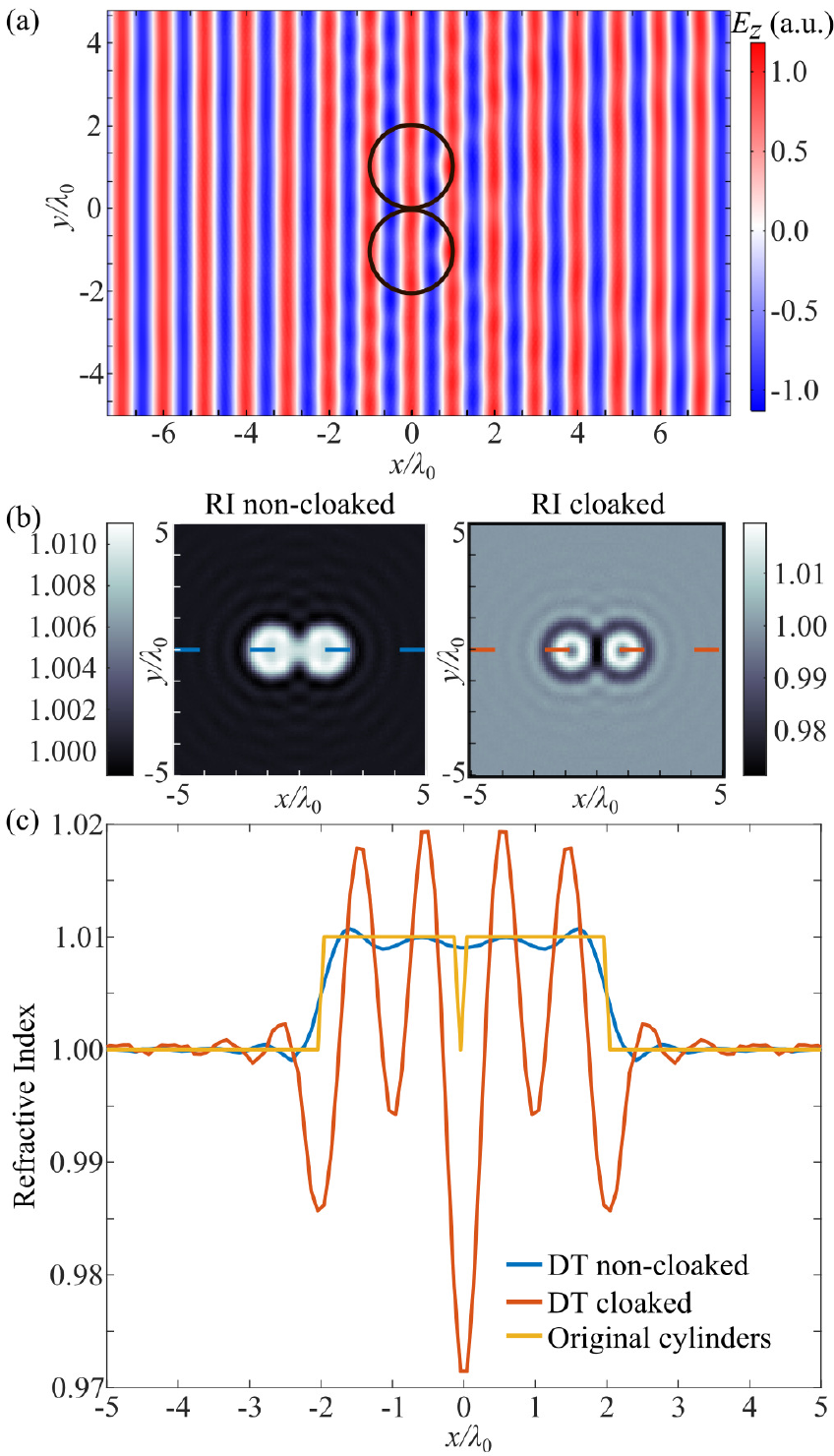}
		\caption{(a) Simulated $z$ component of the electric field resulting from the illumination of two SC-cloaked cylinders ($r=\lambda$) with an $x$-directed plane wave. (b) Tomograms of the non-cloaked and cloaked cylinders obtained via DT (Rytov approximation). (c) Detail of the RI along the dashed lines depicted in (b).}
		\label{FigSC}
\end{figure}

Besides allowing a presence measurement, DT provides information on the apparent shape of the cloaked object. In the tomograms of the studied cylinders, they appear to have a smaller radius and a larger RI variation than their bare counterparts (Fig.~\ref{FigOptb}). Consequently, we ask ourselves whether it would be possible to take advantage of this apparent shrinking, which suggests that covering small dielectric particles with SC coatings may facilitate their differentiation when being closely packed. To study this possibility, we analyze a new configuration consisting of two kissing cloaked cylinders ($r= \lambda$, $n_{\rm{OBJ}}=1.01$), immersed in a vacuum background ($n_{\rm{BG}}=1$) and covered by optimized shells with $\epsilon_{S}=-2$ and $r_{S}=1.0049 \lambda$. The simulation of the corresponding electric field is shown in Fig.~\ref{FigSC}(a). 

	Tomograms of this cylinder arrangement are obtained both with and without SC shells (we use 180 projections). Very similar results are obtained for the Born and Rytov approximations. The tomograms for the latter are shown in Fig.~\ref{FigSC}(b), from which we observe that the bare cylinders are hardly distinguishable, and can be mistaken for a single object [they appear as a cylinder of radius $2 \lambda$ along the central line, as can be seen in Fig.~\ref{FigSC}(c)]. On the contrary, the tomogram of the cloaked cylinders shows two clearly separate profiles, as seen in Fig.~\ref{FigSC}(b), even though they are touching each other, which is a remarkable feature. This fact, combined with the enhanced range between the minimum and maximum RI values, provides a notable advantage in the recognition of closely packed objects, which may find application, for instance, in particle counting or bioimaging.

\section{Transformation optics - Blow-up-a-point cloaks}
A more sophisticated passive cloaking technique that can remove scattering at all orders is based on transformation optics~\cite{Leonhardt2006,Pendry2006}. With this method, a given region of space is hidden by redirecting the illuminating wave around it, avoiding any light absorption or scattering. The rays traversing the cloak bypass the concealed region and turn back to the original path. To study the behavior of this kind of device under DT, we analyze the original 2D cloak proposed in \cite{Schurig2006,Cummer2006}, where a cylindrical region of radius $R_1$ is hidden by a concentric cylindrical shell of radius $R_2$ [see Fig.~\ref{Fig_cloaks}(b)]. This cloak requires the following radius-dependent, anisotropic relative permittivity ($\epsilon$) and permeability ($\mu$) tensor components (in cylindrical coordinates):
\begin{align} 	\label{par_cyl}
	\epsilon_r=\mu_r=\dfrac{r-R_1}{r},\nonumber\\ 
	\epsilon_\phi=\mu_\phi=\dfrac{r}{r-R_1}, \\ 
	\epsilon_z=\mu_z= \left( \dfrac{R_2}{R_2 - R_1} \right) ^2 \dfrac{r-R_1}{r}.\nonumber
\end{align} 
To account for causality and losses, we analyze a realistic version of the cloak in which the components with values below unity ($\epsilon_r$, $\mu_r$, $\epsilon_z$, $\mu_z$) are modeled by Drude and Lorentz dispersive relations, respectively~\cite{Monticone2013,Zhang2008}:
\begin{eqnarray}
	\epsilon_{r{\rm D}}(\mathbf{r})=\epsilon_{r}(\mathbf{r})\left( 2-\dfrac{f_0^2}{f(f+i\gamma_1)}\right), \\
	\epsilon_{z{\rm D}}(\mathbf{r})=\epsilon_{z}(\mathbf{r})\left( 2-\dfrac{f_0^2}{f(f+i\gamma_1)}\right), \\
	\mu_{r{\rm L}}(\mathbf{r})=\mu_{r}(\mathbf{r})\left(1- \dfrac{F}{1+(i\gamma_2/f)-(f_0^2/f^2)} \right), \\
	\mu_{z{\rm L}}(\mathbf{r})=\mu_{z}(\mathbf{r})\left(1- \dfrac{F}{1+(i\gamma_2/f)-(f_0^2/f^2)} \right).
\end{eqnarray}
For the components larger than unity ($\epsilon_\phi$, $\mu_\phi$), we assume a non-dispersive dependence~\cite{Monticone2013,Zhang2008}, given in our case by Eq.~\eqref{par_cyl}. Following Ref.~\cite{Zhang2008}, we take $\gamma_1 = \gamma_2 = \gamma$ and $F = 0.78$. 

\begin{figure}
	\centering
	\includegraphics{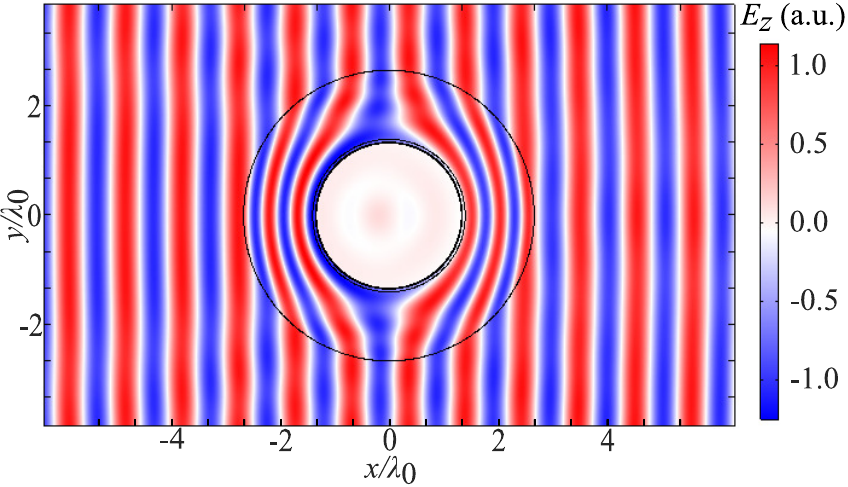}
	\caption{Simulated $z$ component of the electric field resulting from the illumination of a dielectric cylinder covered by a lossless dispersive TO cloak with an $x$-directed plane wave.}
	\label{Fig2aux}
\end{figure}

To validate this TO device, we simulate its response when cloaking a dielectric disk of radius $r_{\rm D} = R_1$ and with a refractive index $n_{\rm{OBJ}}=1.04$. A design frequency $f_0 = 193.55$ THz (corresponding to a wavelength $\lambda_0 = 1.55$ $\mu$m) is chosen, and the dimensions of the cloak are taken to be $R_1 = 2~\mu$m and $R_2 = 2R_1$. The simulation of the electric field at $f_0$ for this configuration is shown in Fig.~\ref{Fig2aux}. 

\begin{figure}
	\centering 
	\includegraphics{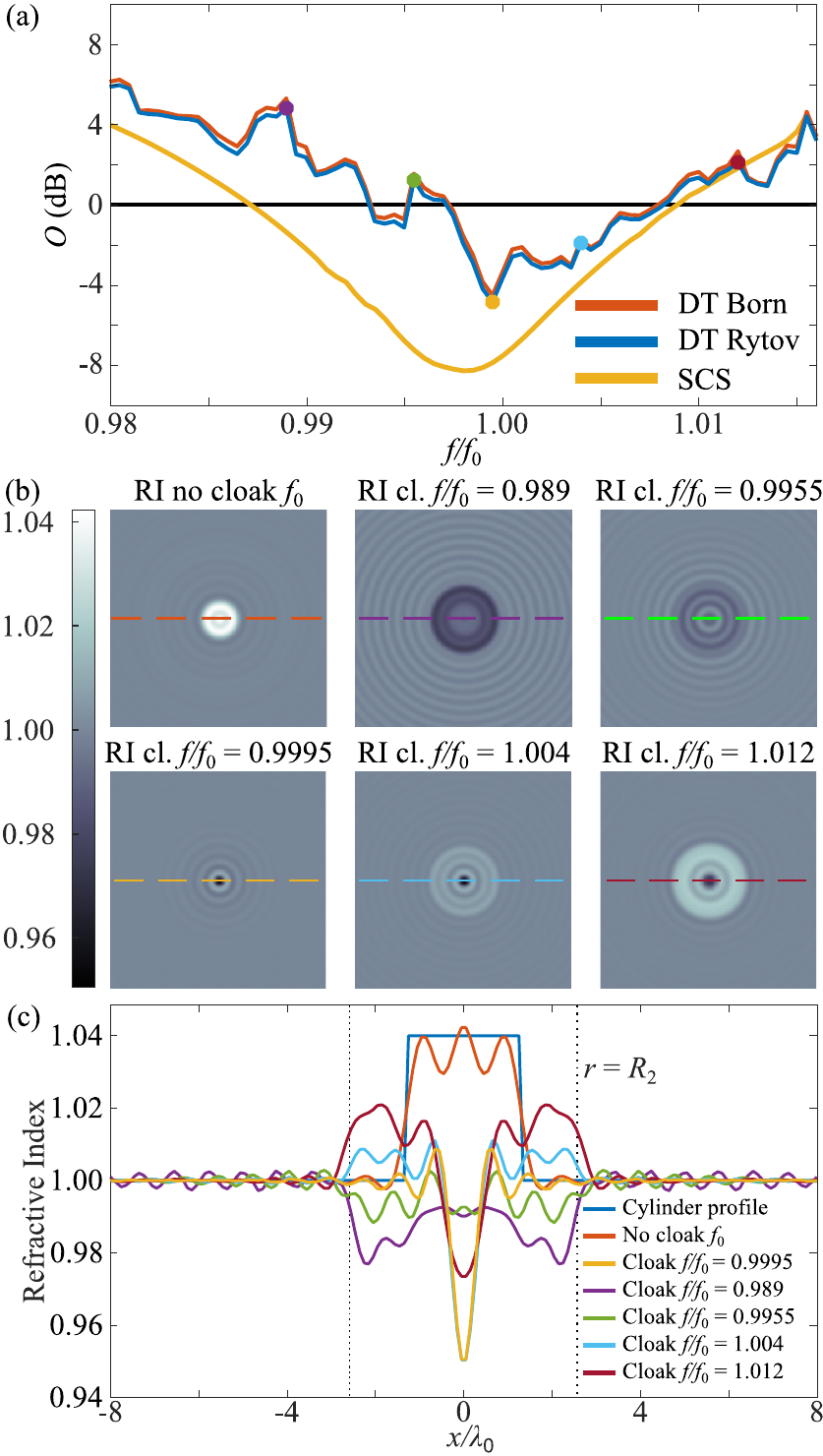}
	\caption{(a) Comparison between the relative observability of a TO-cloaked dielectric cylinder obtained via SCS and DT. (b) Tomograms (Rytov approximation) associated with the non-cloaked cylinder and the cloaked cylinder for different frequencies [corresponding to the color dots in (a)]. The size of the tomograms is $16\lambda_0 \times 16\lambda_0$ (c) Detail of the RI along the dashed lines depicted in (b).}
	\label{FigOT}
\end{figure}

To compare the detection sensibility of the SCS and DT approaches, we study the frequency range from $0.98 f_0$ to $1.016 f_0$, initially considering a lossless material ($\gamma = 0$). We take 1000 projections over a detector with a length $w_{\rm D} = 100 \lambda_0$ and sample the field of each projection at 1000 points. The tomogram of the uncloaked object is depicted in Fig.~\ref{FigOT}(b), which, as seen in Fig.~\ref{FigOT}(c), exhibits the typical ripple that arises when tomography is applied over a large circular object~\cite{Chen1998}. In line with the results reported in \cite{Zhang2008} and \cite{Monticone2013}, Fig.~\ref{FigOT} shows that the curve representing $O_{\rm SCS}$ as a function of frequency has a V shape, with a spectral region around $f_0$ where the SCS of the cloaked configurations is lower than that produced by the bare object. Beyond these limits, the cloaked object scatters more energy than the bare one [values greater than 0 dB in Fig.~\ref{FigOT}(a)]. Although the device was designed to have ideal parameters at $f_0$, the minimum of $O_{\rm SCS}$ is achieved at $f= 0.998 f_0$. This slight change may be due to the numerical error introduced by the simulation software.

The relative observability obtained with DT is also depicted in Fig.~\ref{FigOT}(a), showing a minimum approximately at $f_0$ and the same trend as the $O_{\rm SCS}$ curve. However, the mentioned low observability region found in the SCS analysis is considerably narrowed when using DT for both the Born and the Rytov approximations. Moreover, $O_{\rm DT} > O_{\rm SCS}$ in almost all the analyzed spectrum (especially a frequencies below $f_0$), confirming that DT is more sensitive to lossless realistic cloaks than SCS measurements.\\

To understand this behavior, the Rytov tomograms of the cloaked object at five different frequencies [indicated by colored dots in Fig.~\ref{FigOT}(a)] are shown in Fig.~\ref{FigOT}(b). The RI values of these tomograms along a line passing through the center of the cloak are depicted in Fig.~\ref{FigOT}(c). The Born approximation gives similar results and are not shown. The effect of the cloak in the scattered field at the frequency for which $O_{\rm DT}$ reaches its lowest value [yellow line in Fig.~\ref{FigOT}(c)] gives rise to a tomogram that corresponds to a thinner object than the original one (as in the SC case) and with RI values below $n_{\rm BG}$. As a consequence, while the presence of the cloaked object is quite reduced with respect to that of the uncloaked object, this makes $O_{\rm DT} > O_{\rm SCS}$ at $f_0$. As might be expected, the detection is more precise at frequencies different from $f_0$. In particular, the tomograms associated with the other four frequencies show a variation of the RI spanning a circular region with approximately the same size as the cloak, thus providing information on the shape of the device [see Figs.~\ref{FigOT}(b) and \ref{FigOT}(c)]. In addition to these variations, a concentric ripple of the RI is observed for the frequencies that are farther from $f_0$, i.e., $f=0.989 f_0$ (purple line) and $f=1.012 f_0$ (magenta line), which results in higher observability values. \\

\begin{figure}
	\centering
	\includegraphics{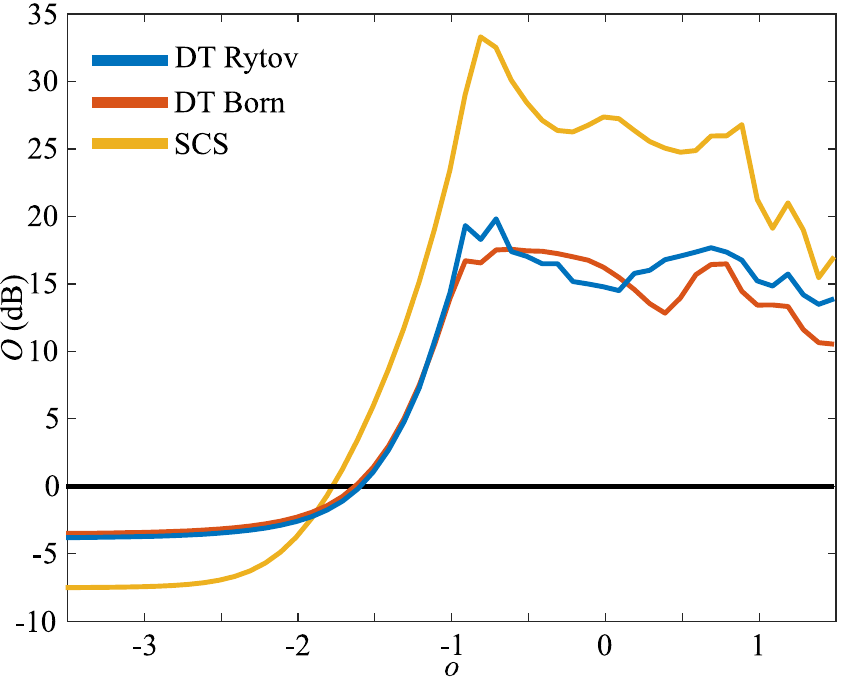}
	\caption{Comparison between the relative observability obtained via SCS and the Born and Rytov approxiamtions of DT as a function of the loss order.}
	\label{FigOTlosses}
\end{figure}

The Drude and Lorentz approximations used to model the cloak also allow us to study the impact of losses in the detection. To this end, we relate $\gamma$ to the design frequency as follows:
\begin{align}
	\gamma=10^{o}f_0,
\end{align}  
where $o$ is termed the loss order. A high value of $o$ will be related to a high value of loss. We sweep the value of $o$ from -3.5 to 1.5 (keeping the frequency constant at $f_0$) and show the corresponding observability in Fig.~\ref{FigOTlosses}. For low losses, the results correspond to the values in Fig.~\ref{FigOT} at $f=f_0$. However, the SCS becomes more sensitive than DT for $o > -1.9$ (note that this corresponds to a material with considerably high loss values, not usually employed to construct cloaks). Interestingly, the maximum values of the observability are achieved around $o = -1$. For $o > 1$, the observability associated with the SCS and DT methods has similar values.

\section{Transformation optics - Polygonal cloaks}
\begin{figure}
	\centering
	\includegraphics{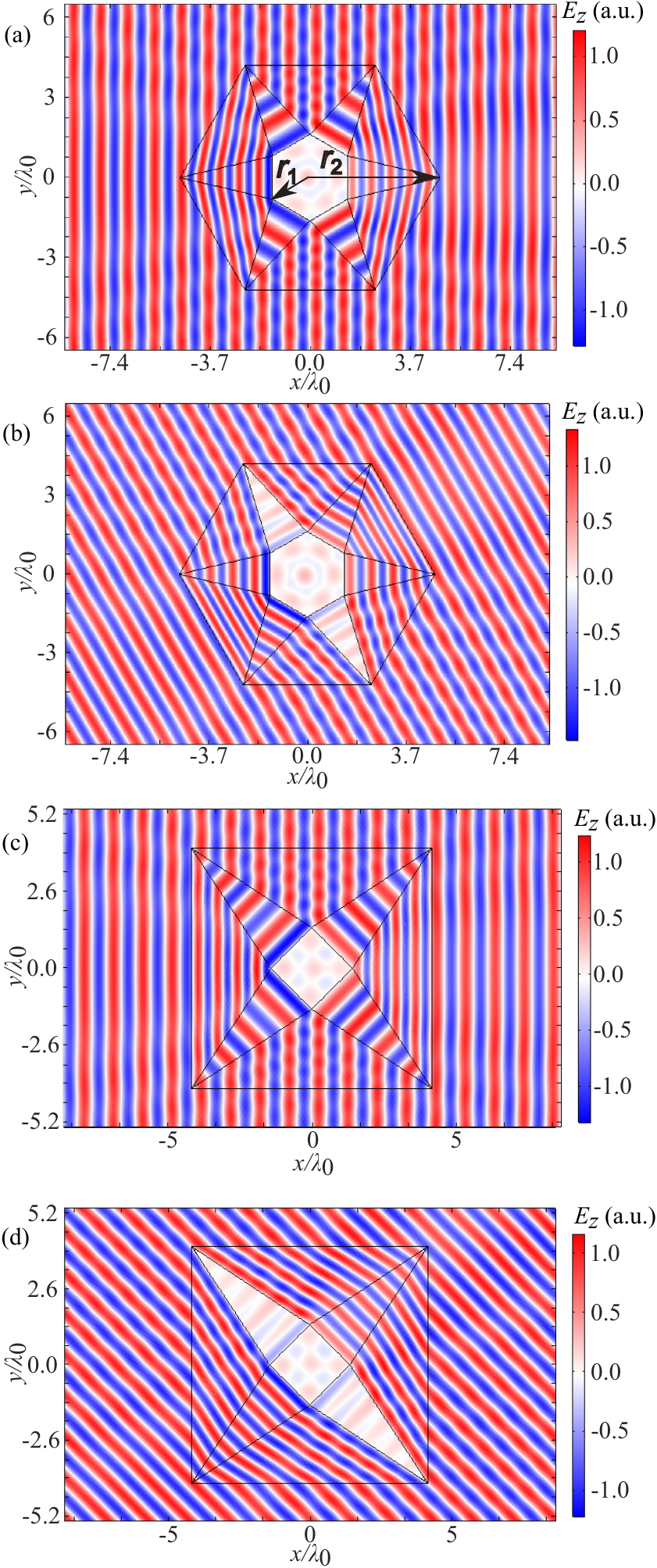}
	\caption{Simulated $z$ component of the electric field resulting from the illumination of different polygonal TO cloaks with an $x$-directed plane wave. (a) Hexagonal cloak illuminated at an angle of 0 rad (b) Hexagonal cloak illuminated at an angle of $\pi/6$ rad. (c) Square cloak illuminated at an angle of 0 rad. (d) Square cloak illuminated at an angle of $\pi/4$ rad.}
	\label{Fig3aux}
\end{figure}
The practical implementation of the ideal TO invisibility cloak studied in the previous section is challenging due to the inhomogeneous character of the $\epsilon$ and $\mu$ tensors, as well as to the extreme values of these tensors near the inner limits of the cloak. One way to address this drawback is to apply a piecewise affine transformation to a polygonal region divided into segments, which results in a homogeneous device with finite constitutive parameters~\cite{Chen2012a}. A simplified version of this cloak can even be implemented with natural birefringent crystals. Moreover, the study of this kind of device allows us to test the DT detection ability for cloaks with non-cylindrical shapes. In particular, here we consider a hexagonal cloak [the details of which are shown in Fig.~\ref{Fig_cloaks}(c) as an example] and a square cloak, following the transformations reported in~\cite{Chen2012a} and~\cite{Orazbayev2016}, respectively. In the case of the hexagonal cloak, the employed dimensions are $r_2=4.8 \lambda$, $r_1=r_2/3$, and $r_0=0.03 \lambda$. The response of this cloak to a TE plane wave is shown in Fig.~\ref{Fig3aux}(a,b) for two different angles of illumination (0 rad and $\pi/6$ rad). As for the square cloak, the inner and outer square boundaries have a side length of $2 \lambda$ and $8.5 \lambda$, respectively. A simulation of its behavior for 0 rad and $\pi/4$ rad illumination angles are shown in Fig.~\ref{Fig3aux}(c,d).

To analyze the behavior of the considered polygonal cloaks under the SCS and DT paradigms, a hexagon and a square with a refractive index $n_{\rm{OBJ}}=1.05$, have been placed inside the cloaks ($n_{\rm BG} = 1$). In the case of the hexagonal cloak, the hidden hexagonal object has a radius $r_1$ and the hidden square has a side length $r_1$. In the case of the square cloak, the hidden hexagon has a radius of $r = 1.02 \lambda$. Through this study, we have verified that, although polygonal invisibility devices work well for any angle of illumination, their performance has a slight angle dependence by construction (the region of virtual space that is expanded is not a point, but a finite domain without cylindrical symmetry). For instance, when the hidden object is the hexagon, $O_{\rm SCS}$ varies from 0.22 in the worst case (angle of illumination of $\pi/6$ rad) to 0.19 when the illumination angle is 0 rad. The SCS ratios for the rest of the cases are gathered in Table~\ref{tab:SCVvsDT}. In all configurations, the SCS is significantly reduced, especially by the square cloak. 

\begin{figure}[h!]
	\centering
	\includegraphics{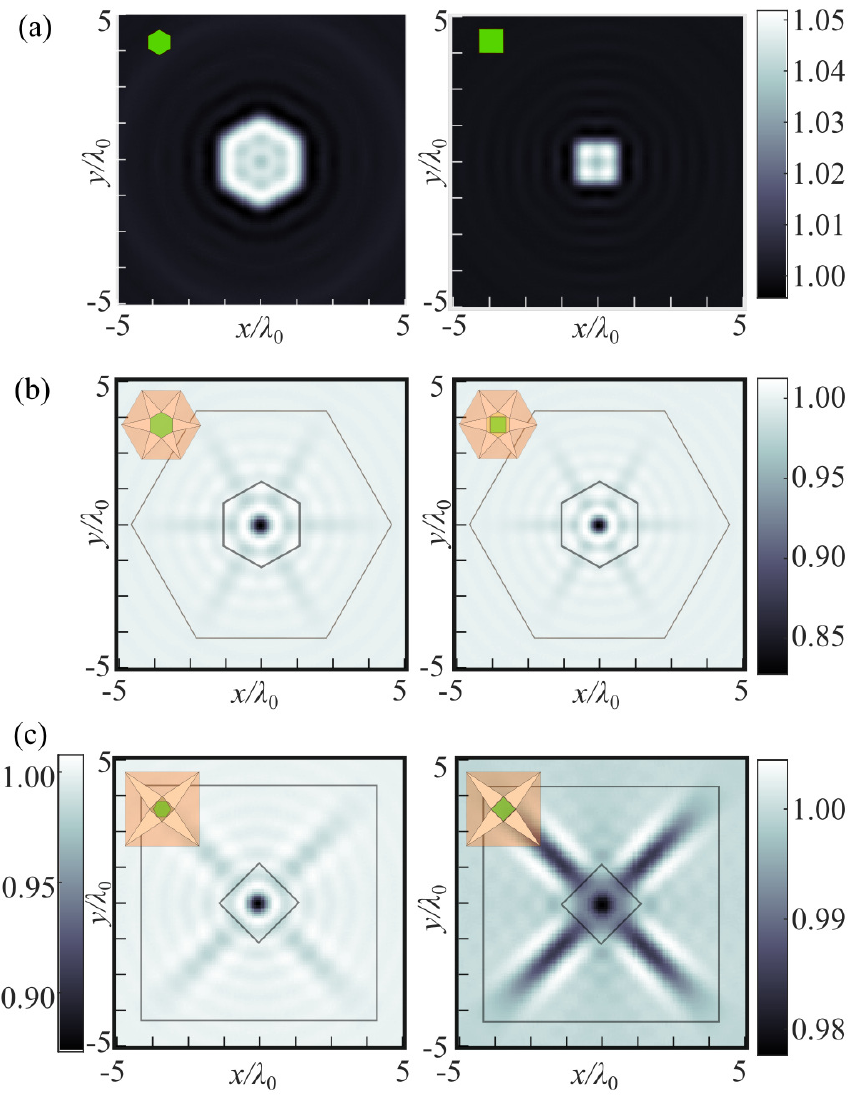}
	\caption{Tomograms (Rytov approximation) of: (a) a bare hexagon and a bare square. (b) Each of these objects hidden within a hexagonal cloak and (c) within a square cloak. The boundaries of the cloaks are depicted as a reference. The schematic insets represent the cloak (pink) and the hidden object (green) for each configuration.}
	\label{FigPC}
\end{figure}

The same analysis is now repeated using DT under the Rytov approximation. The tomograms of the two considered bare objects are shown in Fig.~\ref{FigPC}(a) and are used as a reference. The tomograms of the different cloak-object combinations can be seen in Fig.~\ref{FigPC}(b) and \ref{FigPC}(c). The values of $O_{\rm DT}$ for each case are shown in Table~\ref{tab:SCVvsDT}. As seen, the relative observability associated with the DT approach is always significantly higher than that associated with the SCS method, being higher than unity in some cases (meaning that the cloak enhances the presence of the object in the eyes of tomography).

\begin{table}
	\caption{\label{tab:SCVvsDT} SCS and DT (Rytov) relative observability for each cloak-object combination.}
	\begin{ruledtabular}
		\begin{tabular}{llcc}
			Cloak\footnote{H: Hexagonal, S: Square}&Object&$O_{\rm SCS}$&$O_{\rm DT}$ \\
			\hline
			H & H & 0.22 - 0.19 & 0.74\\
			H & S & 0.52 - 0.47 & 1.69\\
			S & H & 0.40 - 0.18 & 1.51\\
			S & S & 0.13 - 0.05  & 0.81\\
		\end{tabular}
	\end{ruledtabular}
\end{table}

In addition, the tomograms of Fig.~\ref{FigPC} evidence the imaging ability of DT for non-cylindrical cloaks. First, they show a clear correspondence between the size of the retrieved index profile and the outer boundaries of both cloaks. Second, the tomograms possess the same symmetries as the cloaks, as can be seen in Fig.~\ref{FigPC}(b,c). Hence, although the hidden objects are not unmasked by the tomograms, the latter provide important information on the cloaking device, revealing its presence. This means that, while small changes in the scattered energy as a function of the illumination angle do not compromise the cloak performance in terms of SCS (with low values of $O_{\rm SCS}$ well below unity in all cases), they are magnified by DT due to the smart combination of the measured complex field at different angles. Therefore, the high values of $O_{\rm DT}$, together with the ability of DT to determine the shape of the cloak, makes this technique a much better candidate for the identification of this kind of device than the SCS. Furthermore, the most remarkable fact is that this identification has been done at the design frequency of the cloak, i.e., for the exact ideal parameters of the device. It is for this reason that we have not carried out a frequency sweep to analyze the spectral dependence of the observability in this case, although one would expect an increment in the values of this parameter as moving away from $f_0$.

%-------------------------
\section{Conclusions}
%-------------------------
In this work, we have studied the potential offered by DT for the detection of realistic invisibility devices. In this direction, we have shown that this technique is more sensitive than the broadly used SCS for cloaks based on both SC and TO around the design frequency of these devices. Moreover, DT can approximately reveal the shape and the size of all the analyzed devices. 

More specifically, regarding SC cloaks and blow-up-a-point TO cloaks, the presence of the cloaked object has been compared to that of the uncloaked one for different frequencies, finding that the band at which the system possesses low observability according to an SCS meter ($O_{\rm SCS} < 1$) is notably reduced for the DT technique, the observability of the system being always higher or equal under a DT analysis ($O_{\rm DT} \geq O_{\rm SCS}$). In addition, while polygonal TO cloaks have been verified to do a good job in hiding objects as far as the SCS parameter is concerned, DT is able to image the shape of these cloaks and can therefore reveal their presence, compromising their effectiveness. Remarkably, this occurs for the ideal cloak parameters.

Since diffraction tomography represents a more demanding test for the performance evaluation of invisibility cloaks, it might be useful as a new standard in their design and characterization. Additionally, our findings open up a range of potential benefits in a variety of fields, from fundamental advances in invisibility research to bioimaging and warfare applications at the technological level.

Besides, our results indicate that objects cloaked using SC appear to be shrunk from a tomographic point of view. This suggests that the SC technique could be applied to enhance the resolution of DT when imaging small closely-packed particles.

To complete the discussion, we would like to make a note about the robustness of the proposed method. Firstly, although we have employed a high number of projections to obtain the tomograms in this work, a much lower number would be enough in some cases. For instance, the reported polygonal cloak tomograms were obtained using a number of projections ranging from 180 to 360. However, the shape of the cloak can be distinguished even with only a few tens of projections. Secondly, it is worth mentioning that, although the employed DT approach is restricted to isotropic objects possessing a low index gradient, our results prove that this technique also shows a high efficiency when it comes to detecting cloaks with extreme parameters (such as a negative permittivity in the case of SC cloaks, or as anisotropic rapidly-varying materials in the case of TO cloaks), as well as a high index contrast with either the background or the cloaked object (which is the situation in all the analyzed cases).

Finally, given the analogy between optics and acoustics, as well as the advanced state of sonic diffraction tomography~\cite{Gad11,Ozm15}, the proposed technique could be easily extended to the detection of sound cloaks \cite{Nor15,Cum16}.
 
%--------------------------------------------
\section*{Acknowledgements}
F.J. D.-F. acknowledges funding from project TEC2015-63838-C3-1-R OPTONANOSENS (MINECO/FEDER, UE) under grant FPI BES-2016-076818.
%--------------------------------------------

\appendix
\section{Scattering Cross Section (SCS)} \label{AP.SCS}

To evaluate the rate of the electromagnetic energy that is scattered by a particle, the scattering cross-section \cite{Yushanov2013} is defined as:
\begin{align}\label{sigma_sca}
{\rm SCS} = \frac{W_{\rm{SCA}}}{P_{\rm {INC}}}~  ,
\end{align}
where $P_{\rm {INC}}$ is the incident irradiance, defined as energy flux of the incident wave [W/m].  $W_{\rm{SCA}}$ is the scattered energy rate [W], which, in 2D, is derived by contour integration of the Poynting vector associated with the scattered field ($\mathcal{P}_{\rm SCA}$) over a circumference enclosing the particle ($L$):
\begin{equation} \label{sigma_sca_power}
W_{\rm{SCA}}=  \oint\limits_L \mathcal{P}_{\rm SCA} \cdot  \hat{\textbf{n}} ~ dl,
\end{equation}
where $\hat{\mathbf{n}}$ is the unit vector normal to $L$. In the present work, all electromgnetic field and SCS calculations were performed in \textsc{Comsol Multiphysics}.

\end{document}